\documentclass[times,twocolumn,final,authoryear]{elsarticle}
\usepackage{jasr}
\usepackage{framed,multirow}

\usepackage{amssymb}
\usepackage{latexsym}

\usepackage[switch]{lineno}

\usepackage{url}
\usepackage{xcolor}
\definecolor{newcolor}{rgb}{.8,.349,.1}

\usepackage[citebordercolor=white]{hyperref}

\journal{Advances in Space Research}

\begin{document}

%%\verso{Given-name Surname \textit{etal}}
\verso{D. Kang, A. Haungs}

\begin{frontmatter}

\title{The cosmic-ray spectrum in the PeV to EeV energy range}

\author[1]{Donghwa \snm{\textcolor{black}{Kang}}\corref{cor1}}
\cortext[cor1]{Corresponding author:
  donghwa.kang@kit.edu
  %Tel.: +49-721-608-23506
  %fax: +0-000-000-0000;
}
\author[1]{Andreas \snm{\textcolor{black}{Haungs}}}

\address[1]{Karlsruhe Institute of Technology, Institute for Astroparticle Physics, Karlsruhe, 76021, Germany}

\received{15 January 2024}
\finalform{18 June 2024}
\accepted{20 June 2024}
\availableonline{25 June 2024}
\communicated{}

\begin{abstract}
Cosmic rays around the so-called knee in the spectrum at around PeV primary energy are generally galactic in origin. Observations on the form of their energy spectrum and their mass composition are fundamental tools to understand the origin, acceleration and propagation mechanism of high-energy cosmic rays. In addition, it is required to find signatures to clarify the transition from galactic to extragalactic sources, which are believed to be responsible for the highest-energy cosmic rays above EeV.
This brief review focuses on recent experimental results  around the knee of the all-particle energy spectrum and composition in the energy range of the knee up to EeV energies.
\end{abstract}


\end{frontmatter}

%\linenumbers

\section{Introduction}
\label{sec1}
Studying high-energy cosmic rays have been performed to find their sources
since their discovery nearly a century ago.
However, the origin of cosmic rays and their propagation mechanism are not completely understood yet.
The goal of experimental cosmic-ray studies is to determine 
the energy spectrum, mass composition and the arrival direction of cosmic rays.
The energy spectrum of primary cosmic rays reveals some characteristic features,
which hold information about the origin, acceleration and mass composition.

The all-particle energy spectrum of cosmic rays, in general, 
follows a power law ($dN/dE$ $\propto$ $E^{\gamma}$) 
with a spectral index of $\gamma$ around $-3$.
Cosmic rays up to the energy of about 10$^{14}$\ eV can be measured directly by
balloon or satellite experiments, whereas for higher energies direct
measurements cannot provide data with sufficient statistics due to their small
sensitive detection area and exposure time.
Experiments thus have to observe the cosmic rays above 10$^{14}$\ eV indirectly
by measuring extensive air showers \citep{Haungs2003}. 

% knee
Above 10$^{15}$\ eV, the all-particle spectrum has a power-law-like 
behavior with $\gamma \sim -2.7$.
The prominent feature is known as the knee around $3-5 \times 10^{15}$\ eV, 
where the spectral index changes from about $-2.7$ to $-3.1$.
The general explanation of the steepening of the spectrum is 
due to the breakdown of galactic acceleration mechanisms 
of the cosmic rays \citep{Hillas2005}, first of the lowest charges, 
so that the energy positions of the knees for different
cosmic ray primaries would be expected to depend on their atomic number.

% propagation
In addition, propagation of cosmic rays is charge-dependent.
Cosmic rays which undergo faster diffusion will escape more easily from the Galaxy and eventually leads to less flux observed at the Earth. Depending on escaping the sources, the charged particles diffuse in random magnetic field that count for their relatively long confinement time in the Galaxy.
The diffusion model probably works for particles with energies not much larger than $10^{17} \cdot Z$\ eV, where $Z$ is the particle charge.

% acceleration of cosmic rays
Supernova remnants are generally believed to be a source for galactic cosmic rays
from about 10 TeV up to around 1 PeV with acceleration of the particle by the
first order Fermi mechanism \citep{Fermi1949}. The cosmic ray particles gain their
final energy by many interactions each with a small increase of the energy
emitted by supernova explosions.
The maximum attainable energy of charged particles is obtained by 
$E_{knee} \propto Z \cdot B \cdot R$, where $Z$ is the cosmic ray particle charge, 
$B$ and $R$ are the magnetic field strength and the size of the acceleration
region, respectively.

% ankle
In the energy range at around $10^{18}$ eV, 
an important feature is the ankle, which is characterized 
by a flattening of the spectrum.
Cosmic rays above ankle are most probable of extra-galactic origin,
so that in this energy range a breakdown of the heavy component and
a transition from a galactic to an extra-galactic dominated composition 
are expected \citep{Hillas2005, Berezinsky2006, Ginzburg1979}.

Observations performed by various experiments using different measurements techniques
in the energy interval of $10^{15}$ to $10^{18}$ eV showed
the existence of individual knees 
in the spectra of the light, the intermediate and the heavy mass groups of cosmic rays.

In this review, the following experiments presently under consideration are discussed.
KASCADE-Grande \citep{Navarra2004} measured the electromagnetic components with scintillation detectors
and in addition low energy muons of the air showers with shielded scintillators.
IceTop \citep{IceTop2013} uses the Ice-Cherenkov tanks to measure air showers and
has a possibility to include the detection of high-energy muons with IceCube.
The Telescope Array Low-Energy Extension (TALE) \citep{TALE2018} experiment detects low-energy cosmic rays in the PeV energy range using atmospheric fluorescence detectors, which are also sensitive to the directed Cherenkov radiation produced by shower particles.
Their results on the all-particle energy spectrum with mass composition are discussed.
Afterward, a brief discussions on the implication on the results and open questions are followed.

\section{All-particle energy spectrum}
\subsection{Around the knee region}
After the discovery of the knee of cosmic rays, many experimental observations as well as theoretical studies have been performed. However, the origin of the knee remains still controversial.

Most theories plausibly explain
the steepening of the spectrum
by the break of galactic acceleration mechanisms of 
the cosmic rays \citep{Hillas2005} or by the limit 
on their confinement during propagating through the galaxy \citep{Ptuskin1993}.
Another possibility is related to the nature of
hadronic interactions.
In the former models, the knee positions for different cosmic rays depend on their atomic number $Z$, named rigidity dependency, the latter favour spectral changes proportional to the number of nucleons $A$.

Despite, in any of these models, light particles drop out of the spectrum
first, so that the energy position of the knee is expected to vary from light
to heavy elements, i.e. protons would steepen first, then helium, then
CNO. Following that, the heaviest element to be steepened would be the
iron \citep{Bluemer2009, Haungs2003}.

In various experiments, the knee feature at around $3-4 \cdot 10^{15}$\ eV is observed 
in the hadronic, muonic and electromagnetic components
\citep{KASCADE2005, TibetIII2008, GRAPES2021},
as well as in Cherenkov-light measurements \citep{TUNKA2020}
and it is consistent within the experimental uncertainties.
However, a general agreement does not exist yet, namely, on the chemical mass component
for the knee in the spectrum.

\subsection{The iron-knee}
According to the rigidity dependency,
the position of the knee is predicted to vary from light to heavy elements,
so that the iron represented with the heaviest element is expected to be steepened
at around 10$^{17}$\ eV 
following previous KASCADE observations \citep{KASCADE2005}.
Therefore, an understanding of the origin of the steepening 
around $10^{17}$\ eV in the spectrum in terms of mass group separation
is important.

KASCADE-Grande is one of the main experiments for this purpose.
The data of KASCADE-Grande are considered only at energies
$> 10^{16.2}$\ eV to avoid bias effects 
due to different thresholds for different primaries 
at larger zenith angles.
The study is performed subdividing the measured data
in two samples, which are defined as heavy and light mass groups 
based on the correlation between the size of the charged particles ($N_{ch}$) 
and muon numbers ($N_{\mu}$) on an event-by-event basis.
A detailed description on the analysis can be found in Ref.\ \citep{PRL2011}.
The result from the data set of the heavy element group shows a distinct knee-like feature around $10^{17}$\ eV \citep{PRL2011}.
This knee-like feature of the all-particle spectrum as well as of the spectrum of heavy primaries observed by KASCADE-Grande is confirmed by other experiments such as IceTop \citep{PRD2019}, TALE \citep{TALE2018} and the Pierre Auger Observatory \citep{Auger2021}.

\begin{figure*}[t]
  \centering
  \includegraphics[scale=0.37]{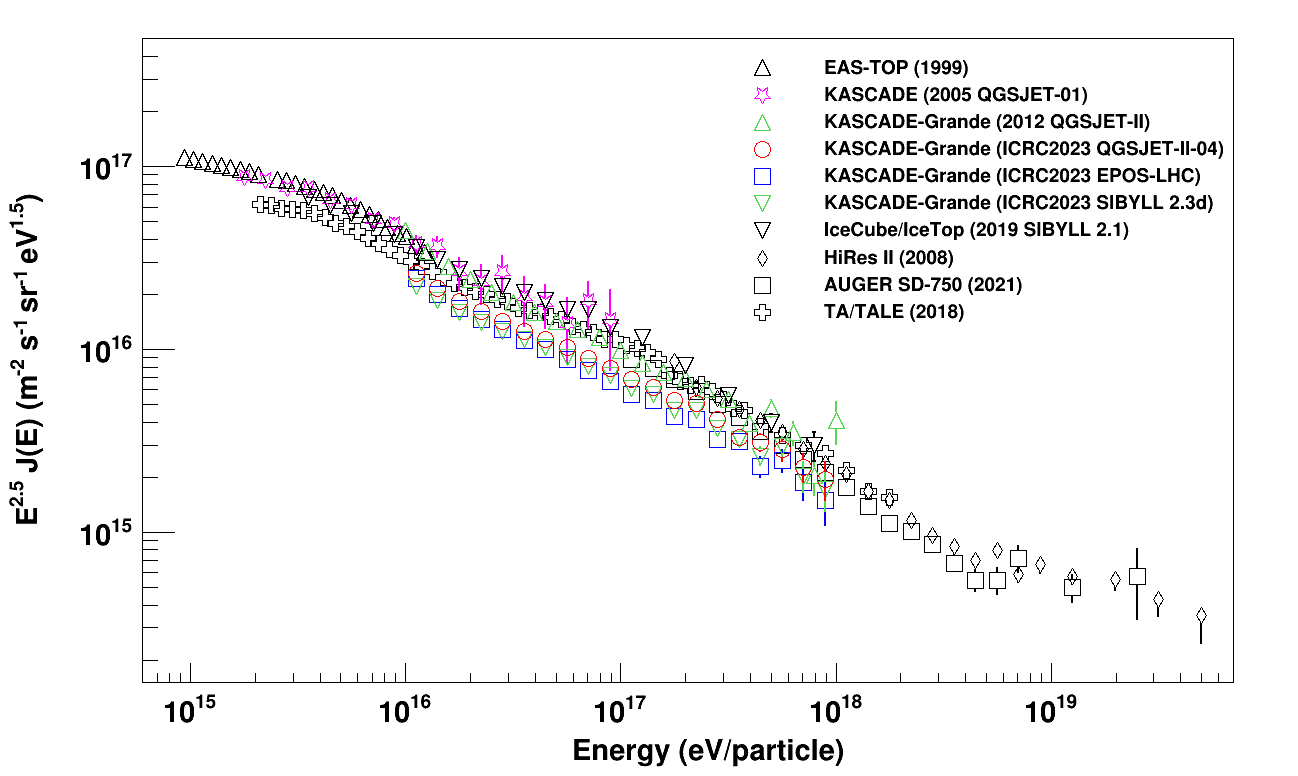}
  \caption{All-particle energy spectrum from various air-shower experiments:
  EAS-TOP \citep{EASTOP1999}, KASCADE \citep{KASCADE2005}, KASCADE-Grande \citep{KASGrande2012}, IceCube/IceTop \citep{PRD2019}, Pierre Auger Observatory \citep{Auger2021}, Telescope Array \citep{TALE2018}.
  } 
\end{figure*}

Figure 1 presents the all-particle energy spectra from various experiments EAS-TOP, KASCADE, KASCADE-Grande, IceCube/IceTop, Pierre Auger Observatory and Telescope Array.
Recent results from KASCADE-Grande \citep{Kang2023} show the spectra based on the different post-LHC interaction models, QGSJET-II-04 \citep{QGSJET2011}, EPOS-LHC \citep{EPOSLHC2015} and SIBYLL 2.3d \citep{SIB23d2020}.
It is interesting to note that the KASCADE-Grande results show relatively small dependencies among the post-LHC hadronic interaction models in use.
However, with regard to the absolute flux, there is for KASCADE-Grande an up to 20\% lower flux compared to the other measurements.
This might be due to the measurements close to sea level, for experiments located at higher altitudes, the variations caused by reconstructions based on different hadronic interaction models seems to be smaller.
Moreover, as KASCADE-Grande measures the total number of electrons and muons, separately, these differences are related to the absolute normalization of the energy scale by the various models.

%IceTop
IceCube/IceTop announced the measurements of all-particle spectrum and energy spectra for different primary mass groups, by combining the events measured by IceTop and the in-ice detector of IceCube in coincidence \citep{PRD2019}.
The resulting spectra derived from three years of IceCube/IceTop data confirmed two features, the knee around 5 PeV and a second knee around 100 PeV.

%Auger
Recently, the Pierre Auger Observatory reported the all-particle energy spectrum
down to around 100 PeV \citep{Auger2021}. This observation suggests that the second knee is not a sharp feature, and this feature is linked to a softening of the spectrum of heavy primaries beginning at around $10^{17}$ eV by the KASCADE-Grande experiment.

%%%
All experiments operate at different observation levels, use different analysis techniques and different hadronic interaction models to interpret their data, nevertheless, a good agreement between the results of different experiments is shown in the energy range of PeV to EeV.
At $\sim10^{18}$\ eV, the KASCADE-Grande result is statistically in agreement with the result of Pierre Auger Observatory.

\subsection{Hardening above the iron-knee}
Following to results from the KASCADE-Grande experiment,
a knee-like feature in the all-particle energy spectrum of cosmic rays
is observed at $10^{16.92}$ eV \citep{PRL2011}.
It is due to the steepening in the flux of heavy primaries.
The combined spectrum of light and intermediate mass components 
was found to be compatible with a simple power law.
However, the spectral feature just above 10$^{17}$ eV
shows a change of the slope, namely, a hardening or ankle-like feature of light primaries.

In KASCADE-Grande, for such a spectral feature, a more detailed investigation
is performed by means of data with higher statistics.
To obtain increased statistics, a larger fiducial area was used 
and the selection criteria for the enhancement of light primaries
is optimized as well. 
Detailed analysis procedures is described in Ref.\ \citep{PRD2013}.
This result is based on the $N_{ch}$-$N_{\mu}$ technique
by calibrating with the hadronic interaction model of QGSJET-II.
A hardening, i.e. an ankle-like feature is visible in the spectrum of the light primaries,
where a change of the spectral slope from -3.25 to -2.79
is observed at an energy of $10^{17.08}$ eV.
This might imply that the transition from galactic to extra-galactic origin
starts already in this energy region.

In astrophysical models,
the transition region from galactic to extra-galactic origin of cosmic
rays is generally expected in the energy range from
10$^{17}$ to 10$^{19}$ eV.
In addition, one should expect a hardening of the proton 
or light primaries components of the cosmic ray spectrum 
to take place below or around 10$^{18}$ eV,
since the onset of the extra-galactic contribution is
dominated by light primaries.

\section{Mass composition}

\begin{figure}[t]
  \includegraphics[scale=0.23]{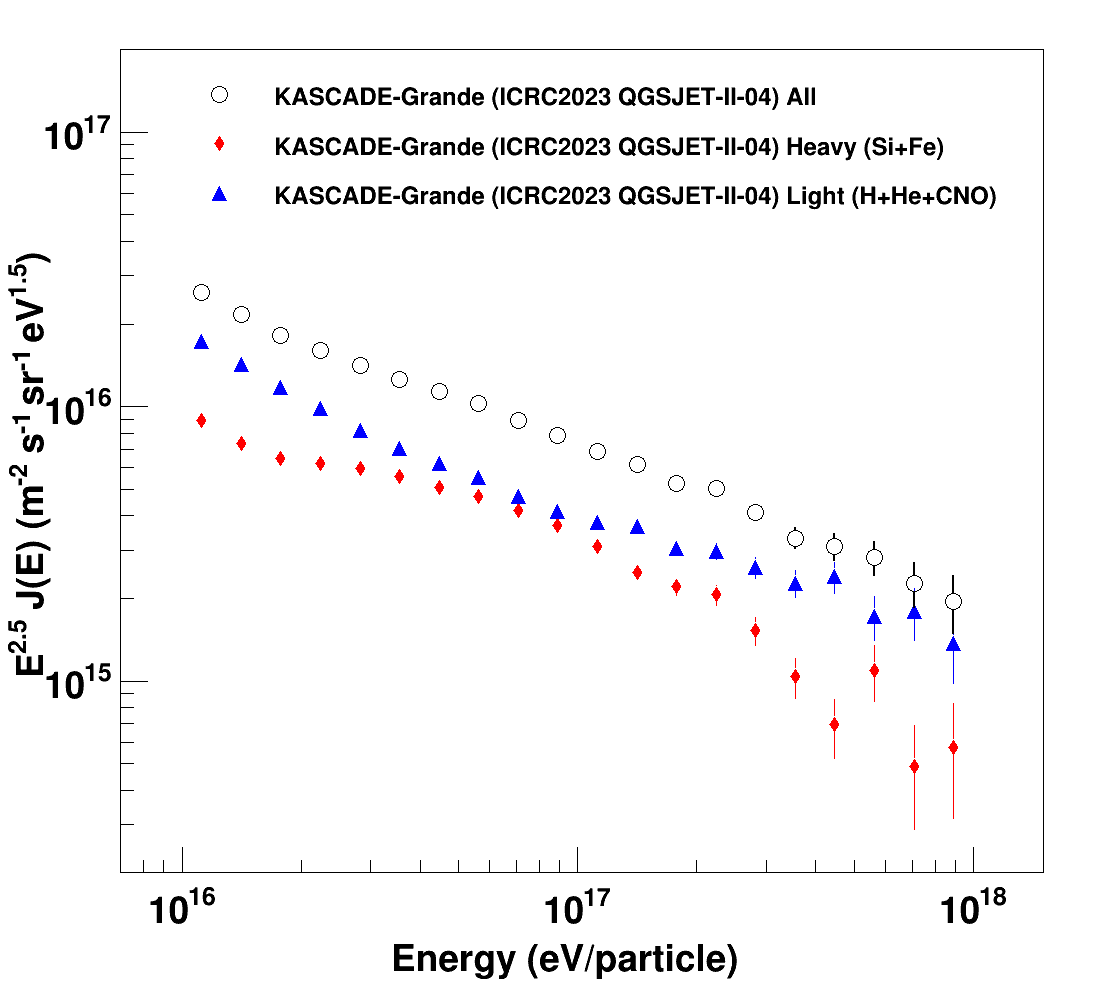}
  \caption{The spectra of heavy (red rhombus) and light (blue triangle) mass groups measured by KASCADE-Grande \citep{Kang2023} are displayed.
  } 
\end{figure}
The mass composition of cosmic rays is one of the most important studies to understand the origin of cosmic rays and the physical processes at the sources.
However, this has to be determined by the measurements of the properties of the atmospheric showers
and has an large influence by the systematic uncertainties related to the hadronic interaction models.
Generally, in ground-based experiments, elemental composition can be investigated in
terms of the individual spectra, e.g. ``light'' and ``heavy'' components or the evolution of $<\rm{ln}A>$ with energy.

Recent results of the energy spectra of heavy (Si+Fe) and light (H+He+CNO) mass groups measured by KASCADE-Grande \citep{Kang2023} are shown in Fig.\ 2.
They are reconstructed by means of the relation $E(N_{ch})$ for two separated samples, using the model-dependent energy calibration function.
The spectrum of heavy components confirms the knee-like structure at around $10^{16.7}$\ eV, where the power spectral index changes from $-2.7$ to $-3.3$.
In the spectrum of light primaries, a hardening feature above about $10^{16.5}$\ eV is also observed and the spectral slope changes smoothly.

The energy spectra of QGSJET-II-04 only is displayed in Fig.\ 2. Details about the individual energy spectra determined on basis of EPOS-LHC and SIBYLL 2.3d can be found in Ref.\ \citep{Kang2023}.
The energy spectra of QGSJET-II-04 and SIBYLL 2.3d show a very similar tendency, whereas the total flux of EPOS-LHC is shifted by about 10\% due to the different ratio of $N_{ch}/N_{\mu}$. In particular, the EPOS-LHC model predicts more muons than the other post-LHC models, so that the data interpretation leads to a more light composition.
In general, the light component sample in KASCADE-Grande is more abundant due to the separation around the CNO mass group. The muon content might also affect some difference of absolute abundances. However, all spectra show similar characteristics of the energy spectrum for all models.

%%IceCube
IceCube presented the mean logarithmic mass $<\rm{ln}A>$, which is derived by the individual fractions from the mass output of the neural network technique. The distribution of the mean mass shows
that composition becomes heavier with increasing energy up to $10^{17}$ eV \citep{PRD2019}.
In addition, the composition spectra from different experiments agree with each other, except the proton spectrum between IceTop and KASCADE. This could be related to the different handling of the intermediate mass groups, which are strongly correlated with other mass components, in particular for the absolute abundance of the primary mass group. 
Another effect could be due to the different observation level.

\section{Astrophysical interpretations}
Various experiments indicate that the knee, the most conspicuous characteristics
in the all-particle energy spectrum is caused mainly by a break
in the spectra for the light primaries,
where the mean mass of cosmic rays increases in this region.
However, the interpretation of the knee of the energy spectrum 
of cosmic rays is still under discussion, in particular, if there are more than one knee for individual primaries, e.g. due to various sources or source populations with different maximum acceleration energies (ARGO \citep{ARGO2015}, LHAASO \citep{LHAASO2023}, GRAPES-3 \citep{GRAPES2023}).   

One of the most general interpretation for the origin of the knee
is that the bulk of cosmic rays is assumed to be accelerated
in the strong shock fronts of SNRs \citep{Fermi1949},
in which the spectrum at the source shows a pronounced break. 
The observed knee produced by the steepening of protons
is at an energy of $E_{knee} \sim 4 \cdot 10^{15}$ eV,
which is possibly close to the size ($\sim$ parsec) of SNRs.

The maximum attainable energy of cosmic ray particles 
has a rigidity dependency \citep{Hillas2005},
so that the energy position of the knees presents a sequence of steepening 
of different nuclei with increasing $Z$.
The steepening of iron is thus easily expected at energy of
$26 \times E_{knee}$.
The first evidence for that has been seen by KASCADE-Grande measurement 
with a knee-like structure of the heavy primary spectrum \citep{PRL2011}.
This result supports the structures of the knee region caused 
as a rigidity dependent feature of the composition.

% Propagation
The acceleration mechanism, i.e. the acceleration of particles 
in $\gamma$-ray bursts is also debated. %\citep{Wick}.
The $\gamma$-ray bursts associated with supernova explosions
are proposed to accelerate cosmic rays from about $10^{14}$\ eV
up to the highest energies.
The propagation effects of the cosmic rays is taken into account 
in this approach, and the knee caused by the leakage of particles 
from the galaxy leads to rigidity dependent behavior.

% Hillas model
Hillas proposed the rigidity dependency of the flux for individual elements.
In this model, the spectra are reconstructed 
with rigidity dependent knee feature at higher energies.
By means of the properties of accelerated cosmic rays in SNRs and
the fluxes derived by KASCADE, Hillas obtained the all-particles flux,
which is not sufficient to describe the measured flux at energy above $10^{16}$\ eV.
For this gap, Hillas proposed a second galactic component, 
which he named as `component B'.
An extra-galactic component becomes significant 
at energies above $10^{19}$ eV.
The flux of galactic cosmic rays extends to higher energies in this case,
therefore, a dominated contribution of the extra-galactic component
is expected only above $10^{18}$ eV.

% Transition between galactic and extragalactic cosmic rays - Berezinsky
The transition between galactic and extra-galactic cosmic rays occurs
most probably at energies around $10^{17}$ and $10^{18}$ eV.
The transition is an important feature since breaks in all-particle 
energy spectrum and in composition are associated with the particle production
mechanism, the source contribution, as well as their propagation. 

In the model of Berezinsky \citep{Berezinsky2006}, 
using the model for extra-galactic ultra-high energy cosmic rays 
and the observed all-particle cosmic ray spectrum by Akeno and AGASA experiments,
the galactic spectrum of iron nuclei in the energy range of
$10^{17}$ - $10^{18}$ eV is calculated.
In the transition region of this model, spectra of only galactic iron nuclei
and of extra-galactic protons are mainly taken into account.
The predicted flux by Berezinsky
at lower energies is well agreeable with results of the KASCADE data.
The transition from galactic to extra-galactic cosmic rays is obviously
seen in spectra of protons and iron nuclei. Above $10^{17.5}$ eV, 
the spectrum can be described by a proton dominated composition.

\section{Summary}
We discussed briefly the all-particle energy spectra 
measured by different ground-based experiments in the PeV to EeV primary energy range.
Several features have been observed in the energy spectrum of cosmic rays:
The first dominant feature, the knee, in the all-particle energy spectrum of cosmic rays 
is a softening of the spectrum at an energy of about $3 \cdot 10^{15}$\ eV, 
which is mainly caused by the light components of cosmic rays.
A knee-like feature in the spectrum of the heavy primaries of cosmic rays,
as well as in the all-particle energy spectrum, is observed
at around $8-15 \cdot 10^{16}$\ eV.
At around $10^{17}$\ eV, an ankle-like structure, i.e. a remarkable hardening,
in the energy spectrum of light components of cosmic rays is observed.
This implies that the transition 
from galactic to extra-galactic origin of cosmic rays might 
occur already in this energy region.

With respect to the mass composition, the ﬁndings of KASCADE-Grande,
IceCube/IceTop and Auger are qualitatively in agreement
with each other.
Though a large uncertainty in the absolute flux/composition,
a common general trend is revealed that composition gets heavier through the knee region and becomes
lighter approaching the ankle.

By means of different hadronic interaction models (post-LHC), there is a shift in the absolute energy scale of the spectra,
but the shape of the spectrum with its structures remains.
Nevertheless, improving the uncertainty in the hadronic interaction models
used to the shower development is expected. 
Up to today the various hadronic interaction models fail to give a consistent picture of reconstructed primary energy and mass of measured extensive air showers for  different observables (measured shower components) and/or different observation altitudes. The most obvious way to solve this problem is to analyse the data from several experiments with the same simulations \citep{Schroder2019}, and to use improved accelerator measurements, for example with a dedicated forward detector \citep{Soldin2023}.  
It would help if the experiments were to make their (also archived) air-shower data freely accessible, as the KASCADE-Grande experiment has done with the KCDC portal \citep{Haungs2018}.

However, in addition, further details are still required to be clariﬁed by future and more sensitive experiments, such as,
whether the origin of the second knee stems from, or
the knee energies of different elements depend on their charge or their mass.
To enable a better understanding of these open questions,
for instance,
in IceCube further analyses are under investigation: more years of experimental data are available and more intermediate elements of cosmic rays are simulated. Furthermore, an investigation of new composition-sensitive parameters is currently under development.
Moreover, an enhancement of the surface array IceTop with a multi-detector array of scintillation detectors and radio antennas is planned \citep{Aartsen2021}.

%\section{Acknowledgments}

%% Bibliography
%% Author year style
\bibliographystyle{jasr-model5-names}
%\biboptions{authoryear}
\bibliography{refs}

\end{document}